\documentstyle[twocolumn]{article}
\def\be{\begin{equation}}
\def\ee{\end{equation}}
\begin{document}
\title{Possible Observation of a Second Kind of Light}
\author{Rainer W. K\"uhne}

\maketitle

\begin{abstract}
According to classical electrodynamics, sunlight that is passed through 
an iron layer can be detected with the naked eye only if the thickness 
of the layer is less than 170nm. However, in an old experiment, August 
Kundt was able to see the sunlight with the naked eye even when it had 
passed an iron layer with thickness greater than 200nm. To explain this 
observation, we propose a second kind of light which was introduced in 
a different context by Abdus Salam. A tabletop experiment can verify 
this possibility. 
\end{abstract}

\vspace{1cm}

August Kundt \cite{Kundt} passed sunlight through red glass, a polarizing 
Nicol, and platinized glass which was covered by an iron layer. The entire 
experimental setup was placed within a magnetic field. With the naked eye, 
Kundt measured the Faraday rotation of the polarization plane generated by 
the transmission of the sunlight through the iron layer. His result was a 
constant maximum rotation of the polarization plane per length of 
$418,000^{\circ}$/cm or $1^{\circ}$ per 23.9nm. He verified this result 
until thicknesses of up to 210nm and rotations of up to $9^{\circ}$. 

In one case, on a very clear day, he observed the penetrating sunlight for 
rotations of up to $12^{\circ}$. Unfortunately, he has not given the 
thickness of this particular iron layer he used. But if his result of a 
constant maximum rotation per length can be applied, then the corresponding 
layer thickness was $\sim 290$nm.

Let us recapitulate some classical electrodynamics to determine the 
behavior of light within iron. The penetration depth of light in a 
conductor is
\be
\delta = \frac{\lambda}{2\pi\gamma},
\ee
where the wavelength in vacuum can be expressed by its frequency 
according to $\lambda = 1/ \sqrt{\nu^2 \varepsilon_0 \mu_0}$. The 
extinction coefficient is
\be
\gamma = \frac{n}{\sqrt{2}}\left[ -1 + \left( 1+ \left( 
\frac{\sigma}{2\pi\nu\varepsilon_0\varepsilon_r} \right)^2 \right) ^{1/2} 
\right] ^{1/2} ,
\ee
where the refractive index is $n=\sqrt{\varepsilon_r \mu_r }$. For 
metals we get the very good approximation
\be
\delta\approx\left( \frac{1}{\pi\mu_0\mu_r\sigma\nu} \right) ^{1/2}.
\ee
The specific resistance of iron is
\be
1/ \sigma = 8.7\times 10^{-8}\Omega\mbox{m},
\ee
its permeability is $\mu_r \geq 1$. For red light of $\lambda =630$nm 
and $\nu =4.8\times 10^{14}$Hz we get the penetration depth
\be
\delta = 6.9\mbox{nm}.
\ee

Only a small fraction of the sunlight can enter the iron layer. Three 
effects have to be considered. (i) The red glass allows the penetration 
of about $\varepsilon_1 \sim 50\% $ of the sunlight only. (ii) Only 
$\varepsilon_2 =2/ \pi \simeq 64\% $ of 
the sunlight can penetrate the polarization filter. (iii) Reflection 
losses at the surface of the iron layer have to be considered. The 
refractive index for electric photon light is given by
\begin{equation}
\bar n^{2} = \frac{n^{2}}{2} \left( 1+ \sqrt{ 1+ \left( 
\frac{\sigma}{2\pi\varepsilon_0 \varepsilon_r \nu} \right)^{2}} \right).
\end{equation}
For metals we get the very good approximation
\begin{equation}
\bar n \simeq \sqrt{ \frac{\mu_r \sigma}{4\pi\varepsilon_0 \nu}}.
\end{equation}
The fraction of the sunlight which is not reflected is
\begin{equation}
\varepsilon_3 = \frac{2}{1+ \bar n}= 
\frac{2}{1+ \sqrt{\mu_r \sigma /(4\pi\varepsilon_0 \nu )}}
\end{equation}
and therefore $\varepsilon_3 \simeq 0.13$ for the system considered. Taken 
together, 
the three effects allow only 
$\varepsilon_1 \varepsilon_2 \varepsilon_3 \sim 4\% $ 
of the sunlight to enter the iron layer. 

The detection limit of the naked eye is $10^{-13}$ times the brightness 
of sunlight provided the light source is pointlike. For an extended 
source the detection limit depends on the integral and the surface 
brightness. The detection limit for a source as extended as the Sun 
(0.5$^{\circ}$ diameter) is $l_d \sim 10^{-12}$ times the brightness of 
sunlight. If 
sunlight is passed through an iron layer (or foil, respectively), then it  
is detectable with the naked eye only if it has passed not more than 
\be
( \ln (1/l_d ) + \ln ( \varepsilon_1 \varepsilon_2 \varepsilon_3 )) \delta 
\sim 170 \mbox{nm}. 
\ee
Reflection losses by haze in the atmosphere further reduce this value. 

Kundt's observation can hardly be explained with classical electrodynamics. 
Air bubbles within the metal layers cannot explain Kundt's observation, 
because air does not generate such a large rotation. Impurities, such 
as glass, which do generate an additional rotation, cannot completely be 
ruled out as the explanation. However, impurities are not a likely 
explanation, because Kundt was able to reproduce his observation by using 
several layers which he examined at various places. 

Quantum effects cannot explain the observation, because they decrease 
the penetration depth, whereas an increment would be required.

The observation may become understandable if Kundt has observed a 
second kind of electromagnetic radiation, the magnetic photon rays. 
They are predicted by a quantum field theoretical model of the 
electromagnetic interaction which includes Dirac magnetic monopoles 
\cite{1}. This model can be constructed in a manifestly covariant 
and symmetrical way if the two potential concept \cite{2,3} is used. 
One potential corresponds to the electric Einstein photon \cite{4}, 
the other one to the magnetic Salam photon \cite{5,6,7,8,9,10}. A few 
years ago, I predicted the interaction cross-section of the magnetic 
photon to be $f=1.5\times 10^{-6}$ times that of an electric photon of 
the same energy \cite{9}. Each process which produces electric photons 
is expected to create also magnetic photons which are $1/f=7\times 10^5$ 
times harder to create, to shield, and to absorb than electric photons. 
Hence, the penetration depth of magnetic photon light of $\lambda =630$nm 
in iron is $\delta /f \approx 5$mm.

To learn whether Kundt has indeed observed magnetic photon rays, his 
experiment has to be repeated.

The easiest test to verify/falsify the magnetic photon is to illuminate a 
metal foil of thickness $1,\ldots ,100\mu$m  by a laser beam (or any other 
bright light source) and to place a detector (charge coupled device or 
photomultiplier tube) behind the foil. If a single foil is used, then the 
expected reflection losses are less than 30\%. If a laser beam of the 
visible light is used, then the absorption losses are less than 15\%. My 
model \cite{9} has to be considered as falsified if the detected intensity is 
less than $1.0\times 10^{-12}$ times the intensity that would be detected 
if the metal foil were removed and the laser beam would directly illuminate 
the detector.

\end{document}